\documentstyle[preprint,aps]{revtex}
\draft

\title{Bridge Hopping on Conducting Polymers in Solution}
\author{Daniel W.~Hone$^{1}$ and Henri Orland$^{1,2}$}
\address{$^1$Institute for Theoretical Physics,
    University of California at Santa Barbara,
    Santa Barbara, CA 93106 \\
    $^2$Service de Physique Th\'eorique, 
    CEA Saclay, 91191 Gif-sur-Yvette, France}
\date{\today}
\narrowtext
\begin{document}
\maketitle
\begin{abstract}
Configurational fluctuations of conducting polymers in solution
can bring into proximity monomers which are distant from each other along 
the backbone. Electrons can hop  between these monomers across 
the ``bridges" so formed.  We show how this can lead to (i) a 
collapse transition for metallic polymers, and (ii) to the observed
dramatic efficiency of acceptor molecules for quenching fluorescence
in semiconducting polymers. 
\end{abstract}

Conducting (conjugated) polymers have been of great interest because
of their unusual electrical and optical properties, combined with
mechanical features very different from those of metallic 
conductors\cite{bredas}. 
In solution, where the polymers are flexible, the 
strong conformational fluctuations can modify both the structural 
and electronic properties\cite{lim,inganas,tubino,annie,hone,honeorland},
which in suitable cases may then be sensitively controlled by 
environmental parameters, such as the temperature.
Theoretical studies have  been limited largely to numerical 
methods.  But in an earlier paper \cite{honeorland} we have demonstrated
the applicability of functional integral methods to these systems,
allowing analytic treatment up to very late stages of
the calculation, with the usual attendant advantages, including the
possibility of treating realistically long chains, which are inaccessible 
in practice to the numerical methods. 

Here we concentrate on the effects of ``bridge conduction", the 
hopping of electrons between monomers distant from one another as
measured along the polymer backbone, but close to each other in
configuration space as a consequence of the bending of the dissolved
polymer. This bridge conduction is believed to be one of the important
mechanisms for electron transfer reactions which play a key role in
several biological processes, such as photosynthesis and cell
metabolism \cite{berini,onuchic}.  A similar phenomenon is expected 
for many chains in solution,
between monomers on different chains which approach one another,
particularly when they are stretched \cite{fyl}.
Bridge hopping was
discussed briefly in a paper by Otto and Vilgis \cite{ottovilgis},
but their results suffer from a critical neglect of the fermion
statistics of the electrons.

We focus here on two characteristic consequences of bridge hopping.
The first, for metallic polymers (partially filled bands)  is
the resultant effective
attraction between remote monomers, and the consequent contribution
to the tendency for collapse from a swollen state.  That collapse
has major consequences for the electronic and optical
properties of the polymer.  The second, for semiconducting polymers,
results from the dramatically enhanced rate for locally excited 
carriers to reach remote parts of the chain, by way of the bridges.
We show that this reasonably  explains the observed\cite{chen} 
ability of a single molecule to quench the fluorescence of a full
chain of hundreds of monomers or more on time scales of picoseconds,
orders of magnitude shorter than the carrier diffusion time along
the chain.

Because our goal is limited to exploring these impacts of bridge
hopping, we have chosen the simplest possible model.  We neglect
completely the dependence of the electronic hopping along the
chain on the local configuration, which dependence was, in fact,
the focus of our earlier paper.  We also take no explicit account
of chain rigidity associated with the moduli of twisting or
bending.  Rather, we make the usual rescaling of the monomer unit
to a Kuhn, or persistence, length, with the new effective units
executing a random self-avoiding walk.

Consider a polymer chain on which a gas of electrons can hop. The polymer is
supposed to be on a d-dimensional lattice with lattice parameter equal
to the Kuhn length $a$. Then the partition function
is the sum over self-avoiding walks for the polymer of the square of an
electronic partition function, $Z_{el}(\{r_n\})$ 
(to account for the spin of the electrons)
, where $r_n$ describes the
position of the $n^{th}$ lattice site visited
\begin{equation}
Z=\sum_{SAW} Z_{el}^2(\{r_n\})
\end{equation}

The electronic partition function can be
written as a functional integral over Fermionic (Grassmanian) 
fields\cite{neg_orl}:

\begin{eqnarray}
Z_{el}(\{r_{n}\}) &=&\int_{c_{n}(\beta
)=-c_{n}(0)}D(c_{n}^{+}(t),c_{n}(t))\exp \left( -\int_{0}^{\beta
}dt\sum_{n}c_{n}^{+}(t)\left( \partial _{t}-\mu \right) c_{n}(t)\right) 
\nonumber \\
&\times &\exp \left( t_{0}\sum_{n}\int_{0}^{\beta
}dt(c_{n}^{+}(t)c_{n+1}(t)+h.c.)+t_{1}\sum_{m,n}\int_{0}^{\beta }dt\Delta
_{r_{m}r_{n}}c_{m}^{+}(t)c_{n}(t)\right)
\end{eqnarray}
where $t_{0}$ is the hopping constant along the chain, $t_{1}$ is the
hopping constant across a bridge, and the operator $\Delta _{rr^{\prime }}$
restricts sites $r$ and $r'$ to be nearest neighbors   on the lattice. 
The antiperiodic
boundary condition accounts for the Fermionic nature of the electrons.
Integration over the Grassman variables yields the result:

\begin{equation}
Z=\sum_{SAW}\exp \left[ 2\mathrm{Tr}\log \left( \left( \partial _{t}-\mu
\right) \delta _{mn}-t_{0}(\delta _{m,n+1}+\delta _{m,n-1})-t_{1}\Delta
_{r_{m}r_{n}}\right) \right] .
\end{equation}

In terms of the Green's function for the electrons along the chain,

\begin{equation}
(\partial _{t}+H-\mu )G_{0}(m,t;n,t^{\prime })=\delta _{mn}\delta
(t-t^{\prime }),
\end{equation}
the partition function can be written as

\begin{equation}
Z=z_{0}^{2}\times \sum _{SAW}\exp \left[ 2{\text{Tr}}\log \left( \delta
_{mn}\delta (t-t^{\prime })-t_{1}\sum _{n^{\prime }}G_{0}(m,t;n^{\prime
},t^{\prime })\Delta _{r_{n^{\prime }}r_{n}}\right) \right] ,
\end{equation}
where $z_{0} $ is the partition function of the gas of electrons free to hop
only along the chain (with hopping constant $t_{0} $):

\begin{equation}
z_{0} = \exp \left( N\int _{-\pi }^{+\pi }\frac{dq}{2\pi }
\log [(1+e^{\beta (\mu+2t_{0}\cos q)}]\right)
\end{equation}

The chemical potential $\mu $ is related to the filling fraction $f $
through the identity:

\begin{equation}
f=\int _{-\pi }^{+\pi }\frac{dq}{2\pi }\frac{1}{1+e^{-\beta (\mu +2t_{0}\cos
q)}}
\end{equation}

Introducing the Fermionic Matsubara frequencies $\omega _{\nu }=(2\nu +1)\pi
/\beta $ where $\nu \in Z $ is a relative integer, we can write the
partition function as:

\begin{equation}  \label{exact}
Z=z_{0}^{2}\times \sum _{SAW}\exp \left[ 2\sum _{\nu \in Z}{\text{Tr}}\log
\left( \delta _{mn}-t_{1}\sum _{n^{\prime }}G_{0}(m,n^{\prime },\omega _{\nu
})\Delta _{r_{n^{\prime }}r_{n}}\right) \right]
\end{equation}
Since $G_0$ is the spatial Fourier transform of $G_0(q,\omega) = 
1 / (i\omega - \mu - 2t_0\cos q)$, we find by direct integration that

\begin{eqnarray}
G_{0}(m,n,\omega ) & = &  \frac{1}{2t_{0}}\frac{z_{0}^{\left| m-n\right| }%
}{\sqrt{\left( \frac{i\omega -\mu }{2t_{0}}\right) ^{2}-1}}  \label{green} \\
z_0  & = & \frac{i\omega -\mu }{2t_{0}} - \sqrt{\left( \frac{i\omega
-\mu }{2t_{0}}\right) ^{2}-1}
\end{eqnarray}
where the square root branch is chosen so that the modulus of $z_0$ is 
less than unity.

Once the Matsubara sums and the trace in the exponent of Eq. (\ref{exact}) 
are done, we are left with a purely configurational partition function with
effective interactions between the monomers. In fact, one can see that  
there exist many-body interactions of all orders.  However, if we assume the 
system not to be at too high a density, we can expect the two-body 
interactions to dominate. We can re-sum {\it to all orders} the 
2-body interactions which arise in the
effective configurational partition function. 
Expanding the exponent in Eq. (\ref{exact}) in powers of $t_{1} $
gives for each positive integer $p$ a term  

\begin{eqnarray}
\text{Tr}\left[ G_{0}\left( m,n^{\prime },\omega _{\nu }\right) \Delta
_{r_{n^{\prime }}r_{n}}\right] ^{p} &=&\sum_{n_{i}}\sum_{m_{i}}G_{0}\left(
n_{1},m_{1},\omega _{\nu }\right) \Delta _{r_{m_{1}}r_{n_{2}}}G_{0}\left(
n_{2},m_{2},\omega _{\nu }\right) \Delta _{r_{m_{2}}r_{n3}} \nonumber \\
&&\cdots G_{0}\left( n_{p},m_{p},\omega _{\nu }\right) \Delta
_{r_{m_{p}}r_{n_{1}}}
\end{eqnarray}
The 2-body contribution comes from the sum of terms such that all 
pairs $\left\{m_{i},n_{i}\right\} $  involve the same two sites
$m$ and $n$:  $\left\{m_{i},n_{i}\right\} $ is either 
$\left\{ m,n\right\} $ or $\left\{n,m\right\} $. That sum is just

\begin{equation}
\left[ G_{0}\left( m,n,\omega _{\nu }\right) +G_{0}\left( n,n,\omega _{\nu
}\right) \right] ^{p}+\left[ G_{0}\left( m,n,\omega _{\nu }\right)
-G_{0}\left( n,n,\omega _{\nu }\right) \right] ^{p}
\end{equation}
Re-summing over all $p$, we find the exact 2-body interaction:

\begin{eqnarray}
\frac{Z}{z_{0}^{2}} \simeq \sum_{all\,\,walks} \exp \Biggl\{ -\frac{g}{2}%
\sum_{m\neq n}\delta (r_{m}-r_{n}) &+& \sum_{\nu =-\infty }^{+\infty}
\sum_{m,n}\bigl\{ \log \left[ 1-t_{1}\left( G_{0}\left( m,n,\omega _{\nu
}\right) +\gamma \left( \omega _{\nu }\right) \right) \right] \nonumber \\
&+&\log \left[
1-t_{1}\left( G_{0}\left( m,n,\omega _{\nu }\right) -\gamma \left( \omega
_{\nu }\right) \right) \right] \bigr\} \Delta _{r_{m}r_{n}}\Biggr\}
\end{eqnarray}
where $\gamma \left( \omega \right) =G_{0}\left( n,n,\omega\right) 
=  \left[2t_{0} \sqrt{\left( \frac{i\omega -\mu }{%
2t_{0}}\right) ^{2}-1}\right]^{-1}$is the equal site Green's function.

Introducing an effective interaction $w $:

\begin{equation}  \label{w}
w(m-n) = -2T\sum _{\nu =-\infty }^{+\infty }  \log \left\{\left[ 1-t_{1}
\left(G_{0}\left( m,n,\omega _{\nu }\right) \right]^2 - 
t_1^2 \gamma ^2\left( \omega _{\nu }\right) \right] \right\} 
\end{equation}
we can write the partition function as

\begin{equation}
Z\simeq z_{0}^{2}\times \sum _{all\, \, walks}\exp \left( -\frac{g}{2}\sum
_{m\neq n}\delta (r_{m}-r_{n})-\frac{\beta }{2}\sum _{m,n}w(m-n)\Delta
_{r_{m}r_{n}}+{3\text{-body}\, \, \text{terms}}\right)
\end{equation}
From Eq. (\ref{green}) we see that $G_{0}(m,n,\omega ) $ decays
exponentially at large separation $|m-n| $, so that the effective
interaction $w(n) $ tends to a finite constant:

\begin{equation}
\lim _{n\rightarrow \infty }w(n) = -2T\sum _{\nu =-\infty }^{+\infty }
\log \left( 1-\frac{t_{1}^{2}}{%
(i\omega _{\nu }-\mu )^{2}-4t_{0}^{2}}\right)
\end{equation}
This quantity turns out to be negative, and thus there is an overall
attraction between distant monomers, due to lateral hopping of the electrons.
To lowest non-vanishing order in $t_{1} $  we have

\begin{equation}
\lim _{n\rightarrow \infty }w(n) 
= -\frac{t_{1}^{2}}{t_{0}}\, \frac{e^{-\beta \mu }\sinh 2\beta t_{0}}{%
\left( 1+e^{-\beta \left( \mu +2t_{0}\right) }\right) \left( 1+e^{-\beta
\left( \mu -2t_{0}\right) }\right) } .
\end{equation}

As a consequence we may expect that, at low enough temperature, the chain
will collapse to a globular state. Indeed, we know from numerical 
studies\cite{theta} that a self-avoiding polymer with  attraction 
of strength $u$ between monomers which are nearest neighbors on
the lattice undergoes a $\theta $-collapse transition at a temperature equal
to $T_{\theta }\simeq 3.64 u $. 
If the long distance part of the
attraction $ w(n) $, which is approximately
constant, dominantly controls the collapse transition, as we might expect, 
then this second order transition should occur at a temperature given by
$T_{c}= 3.64 \, \left| w(\infty )\right| $.
In Fig. 1 we plot $T_{c} $ as a function of the
bridge-hopping parameter $t_{1} $ for various filling fractions, within 
this approximation of constant $w(n)$.  

At shorter distances along the chain, the behavior of $w(n) $ can be
captured by expanding the expression (\ref{w}) for the effective
interaction to lowest order in $t_{1} $. We find

\begin{equation}
w(n)  \approx -4t_{1}\int _{-\pi }^{+\pi }\frac{dq}{2\pi }\, \frac{e^{iqn}}{%
1+e^{-\beta (\mu +2t_{0}\cos q)}} ,
\end{equation}
This  can be calculated numerically, but it is  
instructive to look at its low temperature limit, $\beta t_0 \gg 1$,
for which we
have a simple analytic expression:

\begin{equation}
w(n)=-4t_{1}\frac{\sin \pi fn}{\pi n}
\end{equation}
where $f=1-\frac{1}{\pi }\mathop {\textrm{Arccos}}\left( \frac{\mu }{2t_{0}}%
\right) $ is the filling fraction of the electronic band. Therefore, the 
monomer
interaction oscillates with separation, with wavevector proportional 
to the filling,
and decays as the inverse separation along the chain. In Fig. 2 we plot 
$ w(n) $ as a function of distance $n $ for several values of the
bridge-hopping strength $t_1$ and two temperatures.

Due to these oscillations in $w(n) $  the interaction between monomers can
be attractive or repulsive, depending on their distance along the chain, and
on the filling fraction. This is an effect of the Pauli principle, and may
induce the appearance of modulated and incommensurate phases. We note that
this might preempt the collapse transition discussed above. The search for
such exotic phases is underway and will be presented in a forthcoming
publication.

The simplest method to study the system (at least qualitatively) with the
full variation of $w(n)$ along the chain is to make
a virial expansion, which will converge rapidly at sufficiently low
polymer density.  A straightforward calculation to first order in the Mayer 
function yields:

\begin{equation}
\frac{Z}{z_{0}^{2}Z_{0}}=1+\sum_{m<n}\alpha _{d}\frac{\left( -1+2d(\exp
(-\beta w(m-n))-1)\right) }{(n-m)^{d/2}}+...  , \label{vir2}
\end{equation}
where $\alpha_d$ is a constant of order unity.
As can be seen from this equation, the interaction term is a balance between
the hard-core term $-1$ and the effective interaction due to electron
hopping. As was shown by Flory, a collapse transition occurs when the second
virial term vanishes, that is

\begin{eqnarray}
\sum_{n=1}^{\infty }\frac{1}{n^{d/2}}\Biggl\{ \prod_{\nu =-\infty }^{+\infty
}\left[ 1-t_{1}\left( g\left( n,\omega _{\nu }\right) +\gamma \left( \omega
_{\nu }\right) \right) \right]^{2} &\bigl[& 1-t_{1}\left( g\left( n,\omega
_{\nu }\right) -\gamma \left( \omega _{\nu }\right) \right) \bigr]] ^{2}
\nonumber \\
&-& \frac{2d+1}{2d}\Biggr\} =0
\end{eqnarray}
where $g\left( n,\omega _{\nu }\right) =G_{0}\left( n,0,\omega _{\nu }\right)$. 
The collapse temperature obtained in this way looks very similar to the
results shown in Fig. 1, but the values are systematically higher
by about 50\%.  This is expected in the sense that the low order virial 
expansion leaves out fluctuation effects and always overestimates critical
temperatures.   

The observed quenching\cite{chen}, by localized acceptors, of the fluorescence 
from recombination of distantly created electron-hole pairs requires
ultrafast electron transfer over those long distances (transit times
small compared to the recombination times, of order picoseconds). 
The relevant long distance motion is described by
the electron self-energy at small wavevectors (approximately good quantum
numbers, since the disorder can be averaged on the long length scales
of interest).   In order to calculate this self-energy operator, we must
proceed in the reverse order of the preceding section, namely first trace
out the polymer degrees of freedom so as to obtain an effective electronic
Hamiltonian.  To lowest order in $t_{1}$ the system is in the swollen phase, 
and the relevant configurational ensemble is the set of self-avoiding walks.
Then the effective action for the electrons in first order perturbation
theory gives for the partition function,

\begin{eqnarray}
Z &\approx &Z_{SAW}\int_{c_{n,\sigma }(\beta )=-c_{n,\sigma
}(0)}D(c_{n,\sigma }^{+}(t),c_{n,\sigma }(t))\exp \left( -\int_{0}^{\beta
}dt\sum_{n,\sigma }c_{n,\sigma }^{+}(t)\left( \partial _{t}-\mu \right)
c_{n,\sigma }(t)\right) \nonumber  \\
&\times &\exp \left( t_{0}\sum_{n,\sigma }\int_{0}^{\beta }dt(c_{n,\sigma
}^{+}(t)c_{n+1,\sigma }(t)+h.c.)+t_{1}\sum_{m,n}\int_{0}^{\beta
}dt\,\left\langle \Delta _{r_{m}r_{n}}\right\rangle _{SAW}\,\,c_{m,\sigma
}^{+}(t)c_{n,\sigma }(t)+O(t_{1}^{2})\right) 
\end{eqnarray}
where $\,\left\langle \Delta _{r_{m}r_{n}}\right\rangle _{SAW}$ denotes the
expectation value of the nearest-neighbor operator over self-avoiding walks.
This is a standard quantity in polymer theory\cite{PGG}. It is
related to the probability distribution for the end of the chain through

\begin{equation}
\left\langle \Delta _{r_{m}r_{n}}\right\rangle _{SAW} = P\left(
r=a,m-n\right)   
\sim \frac{A_{d}}{|m-n|^{\nu d +\gamma -1}} , \label{asym}
\end{equation}
where $\nu \approx 3/5$ \ (in $d=3$) is the wandering exponent of a
self-avoiding walk, $\gamma \approx 7/6$ is the susceptibility exponent 
and $A_{d}$ is a non-universal constant of order unity. Note that this
expression is valid for $|m-n|\rightarrow \infty $ . Since this calculation
is done to first order in $t_{1}$ , it does not describe the collapsed
phase, so that if we were interested in the diffusion properties in the
globular phase discussed in the previous section, we would have to take
averages in that phase.  The self-energy of the chain is given by

\begin{eqnarray}
\Sigma (q) &\approx &-2t_{0}\cos q-2t_{1}A_{d}\sum_{n=3}^{\infty }\frac{\cos
(qn)}{n^{\nu d+\gamma -1}}  \nonumber \\
&\sim &B_{d}+2t_{1}A_{d}C_{d}\,q^{\nu d+\gamma -2}+t_{0}\,q^{2}
\end{eqnarray}
where $B_{d}$ is a constant which produces a shift in the chemical potential
of the electrons, and 

\begin{equation}
C_{d} =\int_{0}^{\infty }du\,\frac{1-\cos u}{u^{\nu d+\gamma -1}}  
\approx 1.58\,%
\mathop{\rm in}%
\,3d
\end{equation}

We see that in 3d, the motion of the electrons is super-diffusive, with a
self-energy which behaves as $\Sigma (q)\sim q^{29/30}$ for small $q$ . As a
result, the Einstein law is modified and is replaced by

\begin{equation}
R\sim (t_{1}\,t)^{30/29}
\end{equation}
where $R$ is the radius of the region in which the electrons diffuse during
the time $t$.  For the times of interest in the fluorescence quenching 
problem, of order a picosecond, and for bridge hopping energies of order
0.1 eV this suggests distances of order a few hundred Kuhn lengths, 
enough to explain the observed quenching amplification.  The only other
proposed mechanism of which we are aware is intrachain Forster transfer,
dipolar communication from the created exciton to the chain region near
the quenching center, but we do not know of a quantitative estimate of
the effectiveness of that mechanism.  Our estimates of the bridge hopping
contribution suggest that, if this is dominant, the
quenching efficiency should drop off rapidly for chains an order of
magnitude longer, say, than those of about 1000 monomers which have been
observed to date; we hope such experiments will be undertaken.

We are grateful to Thomas Garel,  Alan Heeger and Andr\'e-Marie Tremblay  
for helpful
conversations.  This work was supported in part by the National Science
Foundation Grant PHY99-07949.

\begin{figure}
\label{fig1}
\caption {The collapse temperature $T_c$ as a function of
the bridge hopping rate $t_1$ for half filling, $f=0.5$ (solid line)
and for $f=0.2$ (dashed line).}
\end{figure}
\begin{figure}
\label{fig2}
\caption {The effective interaction $w(n)$ for half filling,
$f=0.5$ (solid line) and for $f=0.2$ (dashed line). The temperature
is taken to be $T=0.1$, and $t_1=0.1$.}
\end{figure}

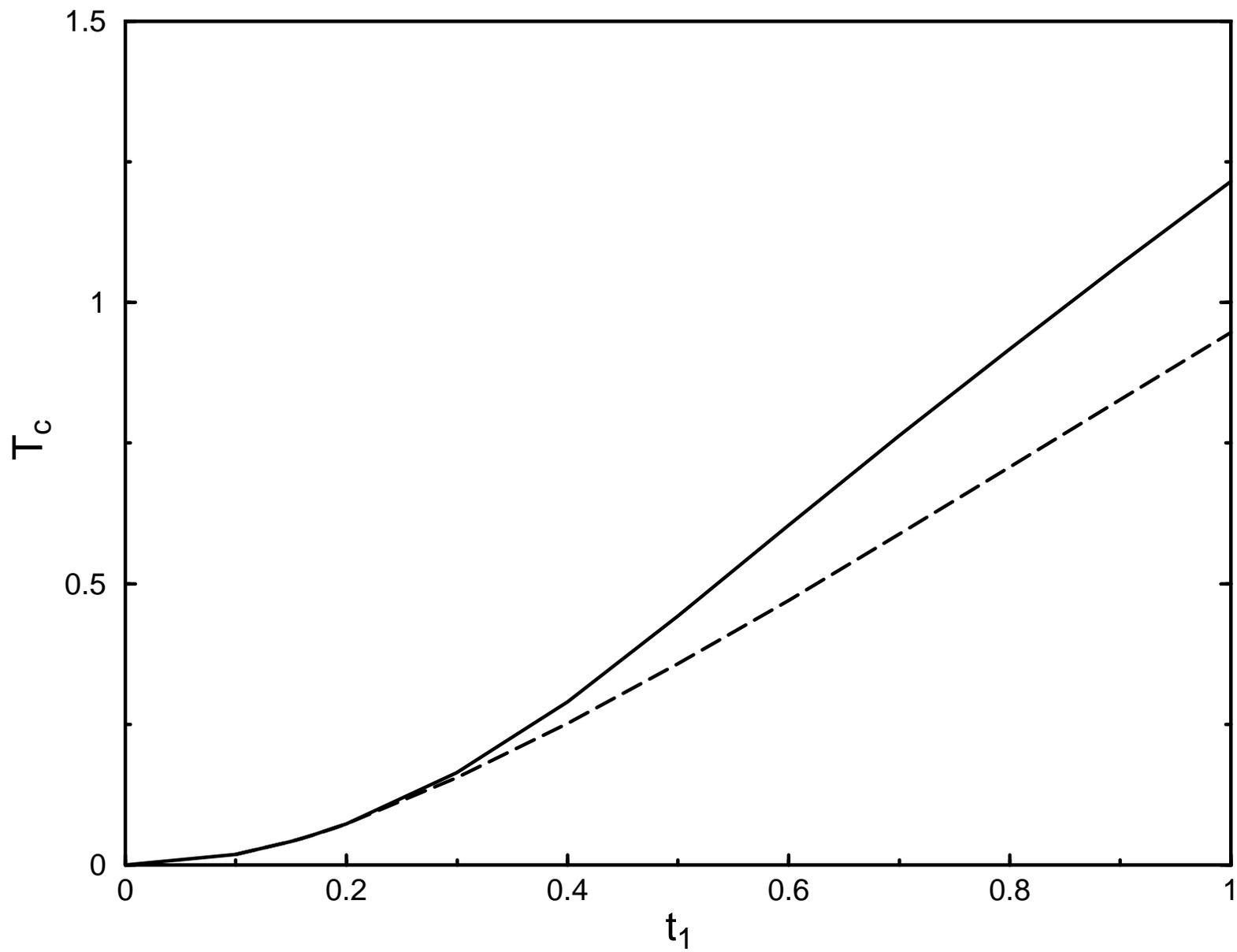
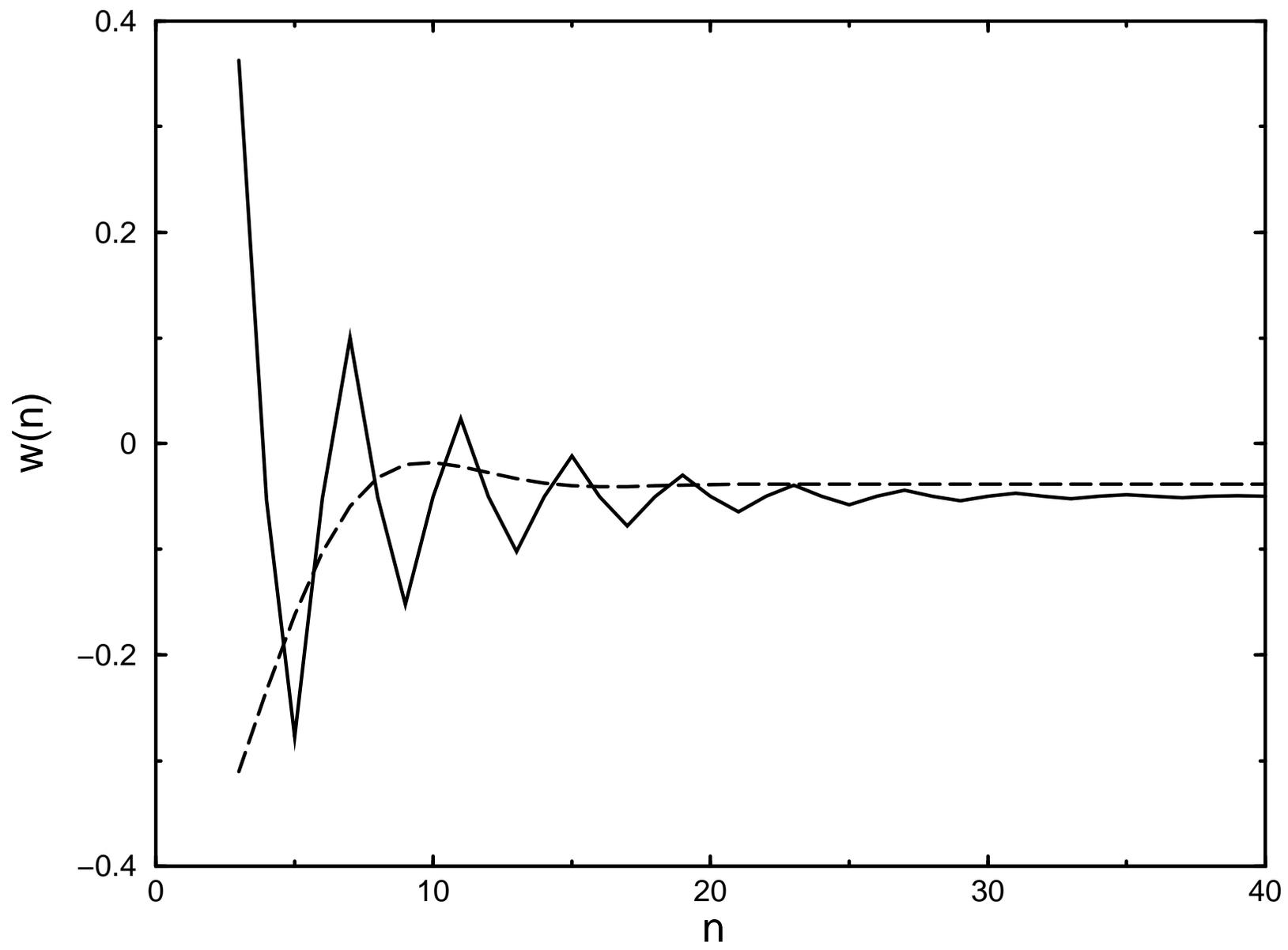
\end{document}